\documentclass[twocolumn,aps,prd,superscriptaddress,nofootinbib]{revtex4-2}

\usepackage{graphicx}
\usepackage{float}
\usepackage{amsmath,amssymb,amsfonts,mathrsfs}
\usepackage[colorlinks,citecolor=blue,linkcolor=blue,urlcolor=blue, breaklinks=true]{hyperref}
\usepackage{breakurl}

\usepackage{caption}
\usepackage{subcaption}

\usepackage{comment}

\usepackage{enumitem} %for enumerations

\usepackage{bbold}
\usepackage{physics} % for \bracket \matrixel \Re \Im
\usepackage[percent]{overpic} %writing over pictures

\usepackage{tikz}
\usepackage{amsmath} % for \text
\usepackage{mathrsfs} % for \mathscr -> scri
\usepackage{xfp} % for fpeval -> floating point expression
\usetikzlibrary{decorations.pathmorphing}

\usepackage[english]{babel}
\makeatletter\AtBeginDocument{\let\@elt\relax}\makeatother % Workaround to avoid a bug due to the babel package. Should be updated soon. See https://tex.stackexchange.com/questions/530439/use-of-x-next-doesnt-match-its-definition-using-memoir-class-toc-chapter#

\usepackage[autostyle, english = british]{csquotes}
\MakeOuterQuote{"}

\usepackage{xcolor}

\begin{document}
	
\title{How the black-to-white hole scenario resolves the information loss paradox}

\author{Pierre Martin-Dussaud}
\email{martindussaud[at]gmail[dot]com}
\affiliation{Basic Research Community for Physics e.V., Mariannenstra\ss e 89, Leipzig, Germany}

\date{ \small\today}

\begin{abstract}
	\noindent This article explains the resolution to the Hawking information loss paradox within the framework of the black-to-white hole scenario.
\end{abstract}
	
\maketitle
	
The information loss paradox can be seen as a no-go theorem regarding the ultimate fate of black holes. Resolving the paradox requires determining which assumptions must be reconsidered. While the black-to-white hole scenario was not originally designed to resolve Hawking's paradox, it presents a plausible trajectory for black holes that circumvents the information loss issue.
	
The long-standing accumulation of confusion surrounding the paradox requires a retelling from its origins. Thus, I will proceed as follows:
\begin{enumerate}
	\item [\ref{sec:hawking}] I explain the paradox and the three primary avenues of resolution;  
	\item [\ref{sec:page}] I discuss the unlikelihood of information emerging in late radiation;
	\item [\ref{sec:remnant}] I introduce the remnant hypothesis;
	\item [\ref{sec:B2W}] I describe the black-to-white hole scenario and explain how it revives the remnant hypothesis.
\end{enumerate}
	
\section{The Hawking information no-go theorem}
\label{sec:hawking}
	
The initial formulation of the paradox appears in Hawking's 1975 paper, \textit{Breakdown of Predictability in Gravitational Collapse} \cite{hawking1976}, submitted less than two years after he demonstrated black hole evaporation \cite{hawking1974,hawking1975}. Hawking introduces this result not as a paradox or a no-go theorem, but as the discovery of a "principle of ignorance," proposing that certain physical processes are fundamentally non-unitary.
	
\begin{figure}[h]
	\centering
	\includegraphics[width = 0.6 \columnwidth]{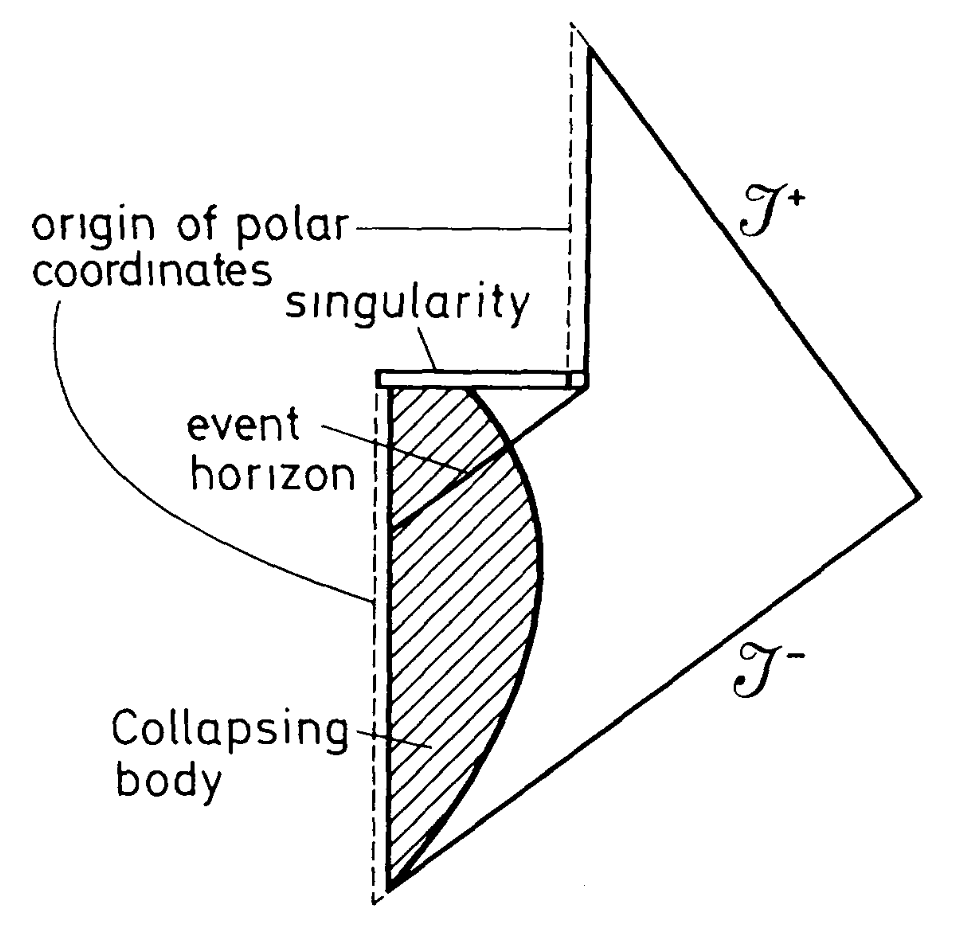}
	\caption[Conformal diagram of an evaporating black hole, as drawn by Hawking in \cite{hawking1975}.]{Conformal diagram of an evaporating black hole, as drawn by Hawking in \cite{hawking1975}.}
	\label{fig:hawking_diagram}
\end{figure}

The argument is best illustrated through the conformal diagram shown in Figure \ref{fig:hawking_diagram}.
In the framework of quantum field theory in curved spacetime, we assume a massless scalar field, $\phi$, propagating on a classical spacetime background. Along $\mathcal{I}^-$, $\phi$ is taken to be in a pure quantum state representing the vacuum. With a Cauchy slicing of spacetime, the state evolves unitarily through the formation of the black hole, progressing from one Cauchy slice to the next. Then, evaporation begins. The purity of the total state remains intact through unitary evolution, but part of the state is hidden by the event horizon, so the state along $\mathcal{I}^+$ is mixed. Overall, the collapse and evaporation processes map a pure state on $\mathcal{I}^-$ to a mixed state on $\mathcal{I}^+$. 
	
\begin{figure}[h]
	\centering
	\includegraphics[width = 0.6 \columnwidth]{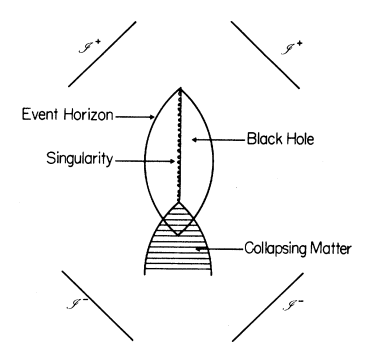}
	\caption{Conformal diagram of an evaporating black hole, as drawn by Hawking in \cite{hawking1976}.}
	\label{fig:hawking-scattering}
\end{figure}

This result may appear puzzling to particle physicists. At the time, particle physics represented the pinnacle of physics, and in this context, all physical processes were seen as S-matrices transforming in-states into out-states in a unitary manner, thus mapping pure states to pure states. Figure \ref{fig:hawking-scattering}, taken from \cite{hawking1976}, illustrates better black hole evaporation from a particle physics perspective.
Hawking demonstrates that evaporation is not a unitary process for an observer at infinity, implicitly raising the question: could incoming particles in a particle accelerator form a tiny black hole that evaporates, with the outgoing state then failing to result from the standard rules of quantum field theory?

Indeed, while quantum mechanics fails to predict the specific outcome of a measurement, it allows to predict the probability distribution over the possible outcomes. By definition, a unitary process linearly evolves the quantum state while mapping one probability distribution to another. Evaporation, however, suggests a physical process in which unitarity is no longer tenable. Hawking argued that physical processes might sometimes require "superscattering operators" that map general density matrices into other density matrices. The resulting mixed out-state holds less information (as measured by von Neumann entropy) than a pure state, implying that some information is "lost."
	
Responses to Hawking’s argument have varied between accepting his conclusion and challenging his assumptions. Three primary positions emerged:
\begin{enumerate}
	\item The conclusion is valid and poses no problem. For instance, Unruh and Wald argue that evaporation's non-unitarity results from discarding part of the system, similar to the standard practice in labs where systems interact with environments \cite{unruh2017}. Here, information "leaks" not into an environment, but into the singularity. Thus, information loss occurs, but it does not contravene any physical principles.
	\item The fixed background metric assumption is invalid since the black hole's mass gradually decreases due to the back-reaction from Hawking quanta emission. This view was explored by Page \cite{page1980}, as detailed in section \ref{sec:page}.
	\item The assumption that the background metric remains classical breaks down when the black hole reaches Planck mass. Quantum gravity effects are then expected to prevent the complete evaporation of the black hole. This view led to the remnant hypothesis (section \ref{sec:remnant}) and includes the black-to-white hole scenario.
	\end{enumerate}
Notice these three positions are not mutually exclusive.
	
\section{Does information emerge in late radiation?}
\label{sec:page}
	
Wald showed in 1975 \cite{wald1975} that Hawking radiation is thermal, implying uncorrelated emitted quanta. In 1980, Page suggested that relaxing the fixed background metric assumption and accounting for back-reaction might yield non-thermal radiation, thus potentially storing "lost information" in correlations among the emitted quanta \cite{page1980}. At the time, this was viewed as the most conservative solution.
	
In 1992, Giddings and Nelson countered this argument with a toy model (two-dimensional dilatonic black holes), estimating back-reaction corrections and finding them too minor to restore the information \cite{giddings1992a}. However, in 1993, Page argued against their findings, showing that Giddings and Nelson’s analysis may not be conclusive \cite{page1993}. More precisely, he argues that the rate of information outflow may initially be too low to show up in the perturbative analysis that Giddings and Nelson had carried out. His model warrants closer examination.
	
Assuming a simplified universe consisting only of a black hole and its outgoing Hawking quanta, we have the total Hilbert space:
\begin{equation}
	\mathcal{H} = \mathcal{H}_{b} \otimes \mathcal{H}_{r}
\end{equation}
where $\mathcal{H}_{b}$ represents the black hole Hilbert space (dimension $b$), and $\mathcal{H}_r$ the radiation Hilbert space (dimension $r$). The latter decomposes as:
\begin{equation}
	\mathcal{H}_r = \underbrace{\mathcal{H} \otimes ... \otimes \mathcal{H}}_{N \ \text{times}}
\end{equation}
with $N$ the number of emitted Hawking quanta and $\mathcal{H}$ the Hilbert space of one quantum. Each time a quantum is emitted, Page assumes that $\mathcal{H}_b$ decreases and $\mathcal{H}_r$ grows by the same amount. Assuming the system is isolated and unitarily evolving, $\mathcal{H}$ remains constant, equaling $b \cdot r$. The number of emitted particle goes as
\begin{equation}\label{eq:radiation-entropy}
	N \propto \log r.
\end{equation}
With $\rho$ representing the overall pure state,
\begin{equation}
	\rho_b \overset{\text{def}}= \Tr_r \rho \quad \text{and} \quad \rho_r \overset{\text{def}}= \Tr_{b} \rho
\end{equation}
denote the black hole and radiation mixed states, respectively. The radiation entropy, $S = - \Tr  \left[ \rho_r \log \rho_r \right]$, also measures entanglement entropy between the radiation and black hole. For maximally mixed $\rho_r$, $S = \log r$. Information content in the radiation can be quantified as the deviation of entropy from its maximum:
\begin{equation}\label{eq:def-information}
	I(\rho_r) \overset{\text{def}}= \log r - S(\rho_r).
\end{equation}
Page wants to compute how $I$ evolves as radiation is emitted (when $r$ grows). In absence of a specific model that would specify $\rho$, the question can be answered on average, over all the possible $\rho$. It happens that, on average, $\rho_r$ and $\rho_b$ are always close to be maximally entangled with each other, and the higher the difference of dimensions between the two subsystems, the more entangled they are likely to be. As a consequence, as long as $r \leq b$, $S \approx \log r$ and $I \approx 0$.

A more accurate computation by Page shows that the initial rate of information outflow goes as
\begin{equation}
	\dv{I}{u} \sim e^{-4 \pi M^2},
\end{equation}
with $u$ the retarded time and $M$ the initial mass of the black hole. This function is not analytic when $M \to \infty$, so that there is no chance to see it via a perturbative expansion at any order, contrary to what Giddings and Nelson had assumed in their analysis.

\begin{figure}
		\centering
		\begin{overpic}[width = 1 \columnwidth]{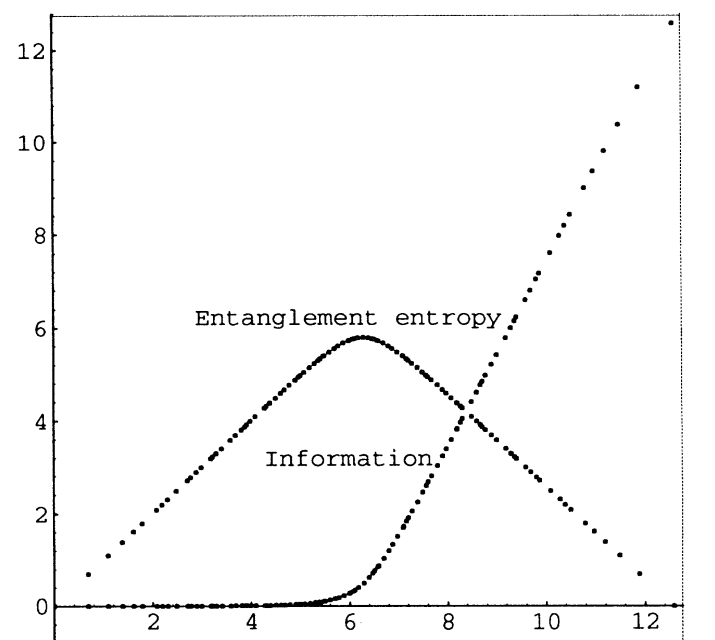}
			\put (40,39) {$S$}
			\put (55,20) {$I$}
		\end{overpic}
		\caption{The radiation entropy $S$ and the information $I$ as functions of $\log r$ (proportional to the number of particles emitted). Taken from \cite{page1993}.}
		\label{fig:page-curve}
	\end{figure}

However, in the second phase of the evaporation, when $r \geq b$, then $S(\rho_r) \approx \log b$, so $I \propto 2 \log r$, so the information outflow is large. The overall behaviour is illustrated in figure \ref{fig:page-curve}.
If Page’s model holds, it implies that "purification" begins once half the particles have been emitted—a moment known as the \textit{Page time}. This result is intriguing because, by this halfway point, the black hole remains macroscopic, meaning the semiclassical approximation used in Hawking’s calculation should still be valid, producing thermal, information-free radiation.

Consequently, it is doubtful that Page’s model accurately represents black hole evaporation. The most contentious aspect of his argument is the assumption that each emitted particle increases the radiation’s Hilbert space dimension while decreasing the black hole’s Hilbert space dimension. Page justified this model by arguing that the dimension of the black hole’s Hilbert space is proportional to the logarithm of the Bekenstein-Hawking entropy, and thus to the black hole’s surface area. As the black hole evaporates and its area shrinks, it naturally follows that the Hilbert space dimension should decrease.
	
However, there are reasons to suspect that Bekenstein-Hawking entropy only accounts for horizon degrees of freedom, omitting interior degrees of freedom \cite{rovelli2017a,rovelli2019}. The black hole’s internal volume actually grows over time \cite{christodoulou2016a}, allowing an old black hole to store more information than the Bekenstein-Hawking entropy alone suggests. This point is crucial to avoid misconceptions: two black holes with the same mass, although similar from the outside, are different from the inside (the older has a larger interior volume). 

While properly accounting for backreaction might still yield non-thermal radiation, this hypothesis remains unconfirmed, and Page  scenario appears unrealistic.

\section{Is information trapped into a remnant?}
\label{sec:remnant}
	
The second controversial assumption in Hawking’s argument is that the black hole evaporates entirely. Extending QFT in curved spacetime to this extreme might exceed its limits, particularly as the black hole approaches the Planck mass $m_P \approx 2 \times 10^{-8} \ \mathrm{kg}$, at which point it is doubtful that a classical Schwarzschild metric remains valid. Although assumptions sometimes extend beyond their conventional domains of validity, in this case, it is plausible that new physics might prevent total evaporation.
	
In his 1975 paper \cite{hawking1976}, Hawking argued that "\textit{because black holes can form when there was no black hole present beforehand, CPT implies that they must also be able to evaporate completely; they cannot stabilize at the Planck mass, as has been suggested by some authors}". However, this reasoning does not hold, as evaporation is not the time-reversal of gravitational collapse. Instead, time symmetry suggests the existence of white holes, not the complete black hole evaporation.
	
In 1982, Hawking presented another argument against remnants, positing that, "\textit{otherwise one would expect the mass density of the universe to be dominated by remnant black holes which would give rise respectively to a very large positive deceleration parameter}" \cite{hawking1982}. Yet, he provides no justification for why remnants should proliferate in such large numbers that they would lead to the observation of a deceleration of the universe expansion.

In 1987, Aharonov, Casher, and Nussinov further examined the Planck-scale remnant concept, which they named the "planckon" \cite{aharonov1987}, proposing that such remnants, if they exist, would be stable. Carlitz and Willey provided an estimate for remnant lifetimes in the same year \cite{carlitz1987}. The following simpler argument was introduced by Preskill \cite{preskill1993}.

During its lifetime, a black hole emits a total number $N$ of quanta. The total amount of radiated energy is given by the initial mass $M$ of the black hole. The average energy $E$ of each quantum is proportional to the temperature $T$ (Wien's displacement law), which is itself proportional to $\frac{1}{M}$ (Hawking's formula). So
\begin{equation}
	N \sim \frac{M}{E} \sim M^2.
\end{equation}
To recover information, a planckian remnant must emit at least $N$ quanta. But this time each quantum has a much smaller energy, because the total energy of the remnant is $1$ (Planck mass), and so each quantum carries an energy $1/M^2$, so that each quantum has a wavelength about $M^2$. Given their correlation with early-time radiation, these quanta would not correlate with each other, requiring emission one by one. This implies an evaporation timescale of at least:
\begin{equation}
	\tau_{\text{remnant}} \sim M^4.
\end{equation}
Thus, the remnant’s lifetime exceeds the expected black hole evaporation time of about $M^3$.
	
The primary objection to the remnant scenario is the issue of infinite pair production rates \cite{giddings1994}. Remnants can be viewed as particles, and for an external observer, they are characterized solely by mass, charge, and spin (in accordance with the no-hair theorem). However, remnants hold substantial information, possessing many internal degrees of freedom and an extensive array of possible orthogonal quantum states. Given the infinite variety of collapse scenarios that could lead to a planckian remnant, such states are infinitely degenerate. Thermodynamically, this would imply that remnants have infinite entropy, making them overwhelmingly favored. From this, some conclude that remnants should appear ubiquitously around us. Yet, a thermodynamically favored state does not necessarily manifest in reality. For example, while the entropy of a book’s ashes-state is much higher than that of its hardcover form, most libraries do not spontaneously combust—fire or another catalyst is required. Typically, systems remain in metastable states, and the most thermodynamically favorable states are only reached over long timescales.

The debate surrounding the remnant hypothesis and the Hawking paradox has evolved considerably since its inception, taking on various nuances. I will not delve into its entire historical development here. However, the previous historical points provide essential context for the emergence of the remnant hypothesis. Now, I will illustrate how quantum gravity has recently reinvigorated interest in this hypothesis.
	
\section{The black-to-white hole resolution}
\label{sec:B2W}
	
Originally, the black-to-white hole scenario emerged not as a solution to Hawking’s puzzle, but as a proposal for black hole fate within background-independent quantum gravity. This scenario naturally suggests a solution to the information paradox through a mechanism akin to remnants.
	
In classical physics, a collapsing star, once within its Schwarzschild radius, must collapse entirely, as matter cannot withstand gravitational compression within a black hole. However, quantum gravity indicates that this view is incomplete, potentially resolving the singularity. Some models propose that a quantum black hole could undergo a tunneling transition to a white hole.
	
White holes are typically defined as either the time-reversal of black holes or as anti-trapped regions. In 1960, Kruskal identified their mathematical existence as part of the maximal Schwarzschild extension \cite{kruskal1960}. In 1964, Novikov first proposed white holes as potentially observable objects \cite{novikov1965}, suggesting they might explain \textit{quasars}, which appeared as point-like sources in distant galaxies, emitting large amounts of energy at radio frequencies\footnote{These signals, first observed in the 1950s, had an unknown origin. Today, quasars are understood to be supermassive black holes whose intense luminosity arises from the extreme heating of their accretion disks.}. Similar ideas were concurrently proposed by Ne'eman \cite{neeman1965}.

The notion of white holes as observable objects was ultimately dismissed about a decade later due to evidence of their instabilities. The first to point this out was Eardley in 1974 \cite{eardley1974}, who noted that even the smallest perturbation in the metric in the distant past could cause a white hole to collapse into a black hole. That same year, Zel’dovich, Novikov, and Starobinskii showed that the first-order quantum correction to the stress-energy tensor diverges along the horizon of a white hole, indicating yet another form of instability \cite{zeldovich1974}.
	
Despite these demonstrations of instability, white holes have been revisited within the context of quantum gravity. The germs of the black-to-white hole scenario appear in a 1979 paper by Frolov and Vilkovisky \cite{frolov1979}. Using an effective model with first-order corrections to the Einstein-Hilbert action, they derive the dynamics of a gravitational collapse. For a spherically symmetric collapse of a null shell of matter, they find that no singularity forms; instead, when the shell reaches \( r=0 \), it reverses and expands outward.

In 2001, Hájíček and Kiefer reached similar conclusions using a reduced quantization method \cite{hajicek2001}. They describe the quantum evolution of a wave packet corresponding to a collapsing null shell of matter and also conclude that the shell undergoes a bounce rather than forming a singularity.

In 2014, Rovelli and Haggard presented an explicit metric describing this transition as observed from outside \cite{haggard2014}. Notably, this metric provides an exact solution to the Einstein equations, except within a small, compact region of spacetime where the equations are violated.

More recently, in 2018, Ashtekar, Olmedo, and Singh analyzed quantum evolution within an eternal black hole \cite{ashtekar2018a}. They foliated the black hole’s interior with spacelike hypersurfaces and approximated quantum dynamics using an effective evolution similar to loop quantum cosmology (LQC). As a result, the classical singularity is replaced by a smooth transition surface, bridging a past trapped region with a future anti-trapped region.

Collectively, these findings support a black-to-white hole transition. However, the interaction between this scenario and Hawking radiation first remained unclear. In \cite{haggard2014}, dimensional analysis suggests that the bounce time observed from infinity would be on the order of \( M^2 \), shorter than the characteristic Hawking evaporation time \( M^3 \). This leads the authors to propose that evaporation can be disregarded in the black-to-white hole scenario.

In contrast, Frolov and Vilkovisky’s earlier work \cite{frolov1979} calculates a bounce time on the order of \( M e^M \), exceeding \( M^3 \). They conclude that evaporation dominates initially, with the bounce occurring only at the final stage.

Loop quantum gravity, in its spinfoam formulation, offers a framework for calculating the bounce time, though it proves challenging without significant approximations. This way, Christodoulou and D’Ambrosio estimate the bounce time as approximately \( M e^{\xi M^2} \) \cite{christodoulou2018a,christodoulou2018b,dambrosio2020}, aligning with Frolov and Vilkovisky’s conclusion that evaporation cannot be neglected.

Thus, in the refined scenario, Hawking evaporation initially dominates, causing the black hole to shrink gradually. Upon reaching the Planck scale, the black hole tunnels into a white hole. This alternative scenario was first examined in \cite{bianchi2018}. In \cite{martin-dussaud2019a}, we further developed and discussed the effective spacetime geometry that arises in the collapse of a null shell of matter. A crucial aspect of this work was to model the backreaction of radiation on the metric.

\begin{figure}[h]
	\centering
	\begin{tikzpicture}[scale=5]
		
		% bounding diagram
		\coordinate (O) at ( 0, 0);
		\coordinate (S) at ( 0,-1); 
		\coordinate (N) at ( 0, 1); 
		\coordinate (E) at ( 1, 0); 
		\coordinate (B) at ( 0.35, 0);
		\draw[thick] (N) -- (E) -- (S) -- cycle;
		
		% labels
		\newcommand{\scri}{\mathscr{I}} 
		\node[right] at (E) {$i^0$};
		\node[above] at (N) {$i^+$};
		\node[below] at (S) {$i^-$};
		\node[above] at (B) {$B$};
		\node[above, rotate=90] at (O) {\small{$r=0$}};
		\node[above right] at (0.5,0.5) {$\scri^+$};
		\node[below right] at (0.5,-0.5) {$\scri^-$};
		
		% matter
		\fill[gray!40] (0.15,0) -- (S) -- (N) -- cycle;
		
		% singularity
		\draw[thick,decorate,decoration={snake, amplitude=0.5mm, segment length=2mm}] 
		(0,0) to[out=0, in=180] (B);
		
		% outgoing radiation
		\foreach \x in {0.07,0.09,...,0.34} {
			\draw[line width=0.2] (\x, 2*\x - 0.7) -- ++(0.85 - 1.5*\x,0.85 - 1.5*\x);
		}
		
		% ingoing radiation into the black hole
		\foreach \x in {0.08,0.09,...,0.28} {
			\draw[line width=0.2] (\x, 2*\x - 0.7) -- ++(-0.61* \x + 0.04, -0.04 + 0.61 * \x);
		}
		
		% ingoing radiation into the white hole
		\foreach \x in {0.28,0.29,...,0.35} {
			\draw[line width=0.2] (\x, 2*\x - 0.7) -- ++(-1.59* \x + 0.32 , - 0.32 +  1.59 * \x);
		}
		
		% black hole horizon
		\draw[thick,cyan] (0,- 0.7) -- (B);
		
		% white hole horizon
		\draw[thick,orange!80!black] (0,0.35) -- (B);
		
		% point at B
		\fill[black] (B) circle (0.02);
		
	\end{tikzpicture}
	\caption{Penrose diagram of an evaporating black-to-white hole.}
	\label{fig:B2W}
\end{figure}
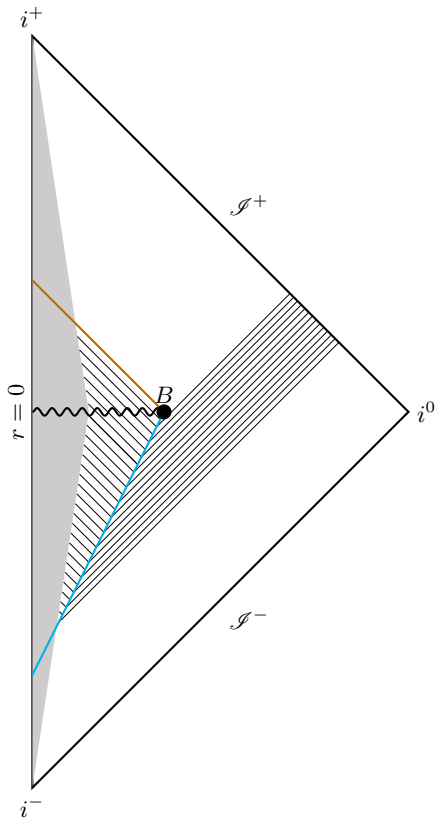

Figure \ref{fig:B2W} illustrates the conformal diagram for this scenario, depicting the collapse of a star.	
The black hole forms through the collapse of a star (shaded area). Pairs of Hawking quanta are generated along the time-like apparent horizon (blue line). In the hatched regions, each line represents either an infalling or an outgoing quantum. These lines demarcate Schwarzschild metrics with progressively decreasing Misner-Sharp mass values, serving as a first-order approximation for the radiation’s backreaction on the metric, as initially introduced by Hiscock \cite{hiscock1981, hiscock1981a}. Radiation falling into the black hole meets the collapsing matter or crosses the would-be singularity, following simple rules as outlined in \cite{dambrosio2018}, and enters an anti-trapped region. Beyond the white hole horizon (orange line), the metric is Schwarzschild with Planck mass. The collapsing matter also crosses the would-be singularity, subsequently bouncing back. Point B represents a compact, Planck-scale region where the classical metric becomes undefined.

This setup enables a resolution of Hawking's paradox within the black-to-white hole framework. The resolution is straightforward, falling within approaches that argue against complete black hole evaporation. Quantum gravity arguments suggest that as the black hole evaporates, it becomes increasingly likely to tunnel into a white hole. Infalling radiation interacts with the star’s matter, creating correlations. Thus, the information carried by outgoing radiation correlates with the bouncing stellar matter. The black hole’s interior is sufficiently large to store these degrees of freedom, even though it connects to the outside only through a Planck-scale horizon. Following the bounce, a Planck-scale white hole remains, persisting for a duration of at least \( M^4 \) before exploding and releasing the stored information\footnote{In a future paper, I will show that the white hole lifetime is $M^5$.}. Although white holes are generally considered unstable and prone to quick decay into black holes, these white holes evade conventional instability arguments because they are Planck-scale \cite{bianchi2018}.

\section{Conclusion}

In the late 1980s, the remnant hypothesis lacked a physical mechanism to would halt black hole evaporation. However, developments in quantum gravity have since argued in favour of a quantum tunnelling phenomenon from black holes to white holes, which would dominate in the final stages of evaporation, thereby giving new life to the remnant hypothesis. This approach offers a straightforward resolution to the Hawking paradox and might even provide insights into the nature of dark matter \cite{rovelli2024}.

\section*{Acknowledgments}

I am grateful to Carlo Rovelli for our discussions on this topic. I am also grateful to Alexandra Elbakyan for her assistance in accessing scientific literature. This work is licensed under CC BY 4.0.

\bibliographystyle{scipost}
\bibliography{hawking-B2W.bib}

\end{document}